\documentclass[sigconf,screen]{acmart}
\usepackage{multirow}
\usepackage{multicol}
\usepackage{graphicx}
\usepackage{subcaption}

\setcopyright{acmlicensed}
\acmPrice{15.00}
\acmDOI{10.1145/3549034.3561176}
\acmYear{2022}
\copyrightyear{2022}
\acmSubmissionID{fsews22maltesquemain-id9265-p}
\acmISBN{978-1-4503-9456-7/22/11}
\acmConference[MaLTeSQuE '22]{Proceedings of the 6th International Workshop on Machine Learning Techniques for Software Quality Evaluation}{November 18, 2022}{Singapore, Singapore}
\acmBooktitle{Proceedings of the 6th International Workshop on Machine Learning Techniques for Software Quality Evaluation (MaLTeSQuE '22), November 18, 2022, Singapore, Singapore}



\title{Are Machine Programming Systems using Right Source-Code Measures to Select Code Repositories?}
\author{Niranjan Hasabnis}
\affiliation{
 \institution{Intel}
 \streetaddress{3600 Juliette Lane}
 \city{Santa Clara}
 \state{California}
 \country{USA}
}
\email{niranjan.hasabnis@intel.com}

\newcommand{\framework}{\texttt{GitRank}}
\begin{document}
\begin{abstract}

Machine programming (MP) is an emerging field at the intersection of
deterministic and probabilistic computing, and it aims to assist software and
hardware engineers, among other applications. Along with powerful compute
resources, MP systems often rely on vast amount of open-source code to learn
interesting properties about code and programming and solve problems in the
areas of debugging, code recommendation, auto-completion, etc.  Unfortunately,
several of the existing MP systems either do not consider quality of code
repositories or use atypical quality measures than those typically used in
software engineering community to select them. As such, impact of quality of
code repositories on the performance of these systems needs to be studied.

In this preliminary paper, we evaluate impact of different quality repositories
on the performance of a candidate MP system. Towards that objective, we develop
a framework, named {\framework}, to rank open-source repositories on quality,
maintainability, and popularity by leveraging existing research on this topic.
We then apply {\framework} to evaluate correlation between the quality measures
used by the candidate MP system and the quality measures used by our framework.
Our preliminary results reveal some correlation between the quality measures
used in {\framework} and ControlFlag's performance, suggesting that some of the
measures used in {\framework} are applicable to ControlFlag. But it also raises
questions around right quality measures for code repositories used in MP
systems. We believe that our findings also generate interesting insights towards
code quality measures that affect performance of MP systems.

\end{abstract}

\maketitle

\section{Introduction}

Last several years in the field of software engineering have seen emergence of
new types of software systems that employ techniques from combination of
artificial intelligence (AI), deep learning (DL), machine learning (ML), and
formal methods to assist software developers. Gottschlich et
al.~\cite{gottschlich:2018:mapl} have coined the term \emph{Machine Programming}
(MP) to describe these systems.
In general, MP systems aim to improve programmer productivity by assisting them
in various tasks such as automatic code generation~\cite{hasabnis:2016:eissec,
hasabnis:2016:lisc, kamil:2016:pldi}, code recommendation and
search~\cite{luan:2019:aroma, ye:2020:misim}, automated bug detection and
program repair~\cite{dinella:2020:hoppity, hasabnis:2020:controlflag,
Yasunaga:2020:drrepair}, automatic completion of program constructs for
integrated development environments (IDEs)~\cite{chen:2021:codex,
gao:2020:OOPSLA, Svyatkovskiy:2019:KDD}, and language-to-language
translation~\cite{roziere:2020:transcoder}. Several commercial tools, such as
GitHub CoPilot~\cite{github_copilot} and OpenAI Codex~\cite{chen:2021:codex,
openai_codex}, can also be classified as MP systems. GitHub CoPilot is an
AI-based pair programming system that suggests completions of source code.
OpenAI Codex is an AI system that learns to generate Python
programs that solve problems specified in natural language descriptions.

Majority of the existing MP systems rely on data mined from open-source code
repositories to train their AI models. This is either because these systems use
self-supervised or unsupervised learning techniques to mitigate the shortage of
labeled code datasets\footnote{IBM's project CodeNet~\cite{puri:2021:codenet}
aims to alleviate the problem of shortage of labeled code datasets by
providing curated code examples.} (required for supervised learning techniques) or
for cost purpose (as labeling is an effort-intensive task). For instance, Codex
is trained on the dataset containing functions written in Python and their
docstrings (documentation), and the dataset is obtained by mining 54 million open-source
repositories on GitHub.

Unfortunately, the reliance of MP systems on open-source code repositories can
be problematic, if quality of the open-source code repositories used for mining
is left unchecked. And this is primararily the case with most of the MP systems.
For instance, ControlFlag~\cite{hasabnis:2020:controlflag} considers open-source GitHub repositories having at
least 100 stars as good quality and mines programming patterns from those
repositories to generate a dataset of known, good patterns for training.
However, it is unclear if the number of stars of a GitHub repository indeed
reflects its actual quality, when commonly-used quality models, such as ISO/IEC
25000:5000~\cite{iso25000}, consider other quality measures such as accuracy,
functionality, etc, and also when the number of GitHub stars is typically
considered a popularity metric than a quality
metric~\cite{hudson:2016:github_stars, hudson:2018:github_stars}.  As such, it
may be plausible to launch an attack on ControlFlag and degrade its performance
by intentionally increasing the number of stars of a bad-quality repository!
Quality of the training
dataset is one of the important factors in determining quality of the learned AI
model. As such, impact of quality of code repositories on the performance of MP
systems needs to be studied.


In this paper, we evaluate if quality of open-source repositories indeed impacts
the performance of the MP systems by choosing ControlFlag as a candidate for
evaluation. Towards that goal, we develop a framework, named {\framework}, to
rank open-source repositories on quality, maintainability, and
popularity\footnote{We could not find existing open-source framework that
combines metrics of our interest and ranks repositories.}. We derive quality
measures from existing literature~\cite{Jarczyk:2014, ludwig:2017,
samoladas:2008:sqooss, stamelos:2002:codequality, spinellis:2009:sqooss1} and
known quality models such as ISO/IEC 25000:5000~\cite{iso25000},
SQO-OSS~\cite{samoladas:2008:sqooss}.  We also consider commonly-used popularity
and maintainability measures~\cite{Chatzidimitriou:2018:npm_miner, npmregistry}
as existing research suggests that the number of GitHub stars --- the measure
used by ControlFlag to determine quality of a repository --- is a popularity
measure~\cite{hudson:2016:github_stars, hudson:2018:github_stars}.
We then apply {\framework} to rank randomly-selected 500 GitHub repositories,
having C++ as their primary language and at least 100 GitHub stars --- the same
criteria used by ControlFlag. We then use the ranked repositories to analyze
their impact on the performance of ControlFlag. Our findings so far suggest that
ControlFlag's performance shows some correlation with the source code measures
used in {\framework}. The performance, on the other hand, shows negative
correlation with the code quality measures used in ControlFlag.  We believe that
our findings provide first set of insights into the types of MP systems that
could be robust to quality of repositories.

\textbf{Contributions.} This paper makes following contributions:
\begin{itemize}

\item We believe to the best of our knowledge that this is the first attempt at
evaluating the impact of quality of the code repositories on an MP system,
specifically ControlFlag.

\item We gather insights from existing research in the field of software
engineering to measure quality, maintainability, and popularity of code
repositories and build {\framework} that combines several of the
commonly-used metrics together.
We, nonetheless, do not claim the novelty of the framework. All the
measures that we use to rank repositories are known.

\end{itemize}



\section{Framework for Ranking Open-source Repositories}
\label{section:framework}

We now describe our framework, named {\framework}, to rank open-source
repositories based on quality, maintainability, and popularity. (We discuss
{\framework} here briefly to provide enough context for this work ---
reference~\cite{hasabnis:2022:gitrank} describes the framework in full details.)
{\framework}  can be broken down into two phases. Given a set of open-source
repositories to be ranked, in the first phase, we obtain values of quality,
maintainability, and popularity measures of every repository individually. In
the second phase, we then compare values of measures across repositories to
calculate quality, popularity, and maintainability scores that are then used to
calculate overall score for every repository.

\paragraph{\textbf{Phase 1: Obtaining values of measures for every repository}}

Let us denote the set of open-source repositories to be ranked as $R$, and the
set of quality, maintainability, and popularity measures that we use as $M$. The
outcome of the first phase is a 2-dimensional table of $R{\times}M$, where every
row in the table corresponds to a repository from $R$, while every column
corresponds to one of $M$ measures. Table~\ref{table:measures} contains the set
of measures that we use.  Out of a typical set of source code measures used in
software engineering, we consider a subset that is applicable to open-source
repositories and is common across set of languages. For instance, we do not
consider depth of inheritance, a common measure used in object-oriented
languages to determine complexity of maintaining code, as it is not applicable
to other languages that are not object-oriented. Note, however, that the
selection of code measures currently is ad-hoc --- we selected a set of commonly
used measures to study their applicability to MP systems\footnote{This
is predominantly given the preliminary nature of the study. We have nonetheless
surveyed several software engineering topics to select the code measures.}. We
explain some of the non-obvious measures below.

\begin{table}
\begin{center}
\begin{footnotesize}
\begin{tabular}{p{1cm}|l|l}
\hline \hline
\textbf{Category} & \textbf{ID} & \textbf{Description of the measure} \\
\hline \hline
\multirow{5}{*}{Quality} & \texttt{cc} & Average Cyclomatic complexity of a repository \\
 & \texttt{sty} & Number of style errors per LoC \\
 & \texttt{sl} & Number of security errors of low severity per SLoC \\
 & \texttt{sm} & Number of security errors of medium severity per SLoC \\
 & \texttt{sh} & Number of security errors of high severity per SLoC \\
\hline
\multirow{6}{1cm}{Maintain-ability} & \texttt{mi} & Average maintainability index of a repository \\
 & \texttt{c2y} & Number of closed issues and pull requests over last 2 years \\
 & \texttt{c1y} & Number of closed issues and pull requests over last 1 year \\
 & \texttt{c6m} & Number of closed issues and pull requests over last 6 months \\
 & \texttt{c1m} & Number of closed issues and pull requests over last 1 month \\
 & \texttt{cm} & Number of commits per day \\
\hline
\multirow{3}{*}{Popularity} & \texttt{ss} & Number of subscribers per day \\
 & \texttt{str} & Number of stargazers per day \\
 & \texttt{fr} & Number of forks per day \\
\hline
\end{tabular}
\end{footnotesize}
\end{center}
\caption{Measures and their codes used in {\framework}}
\label{table:measures}
\vspace{-0.12in}
\end{table}

\begin{itemize}

\item Cyclomatic complexity of a source code is a well-known measure of the
structural code complexity~\cite{mccabe:1976:cc}. We determine the
cyclomatic complexity of a repository by averaging over cyclomatic complexity
of individual functions.

\item We consider code formatting as a code quality measure. We divide the
total number of formatting errors by the lines of source code
(SLoC)~\cite{Nguyen:2007:sloc} to obtain the density of style errors.

\item Security issues are an important class of errors that need no explanation.
An example of a security error would be using a potentially dangerous function
such as \texttt{strcpy} in C language (CWE-676~\cite{cwe676}).  Security issues
could be of varying severity levels. In {\framework}, we consider three severity
levels: low, medium, and high. We divide the total number of security errors
reported for every level by the lines of source code (SLoC) to obtain the
level-specific error density.

\item Maintainability index is a well-known, composite metric that incorporates
a number of traditional source code metrics into a single number that indicates
relative ease of maintaining source code~\cite{oman:1992:mi}. We use the
modified formula of MI~\cite{welker:1997:modified_mi} to obtain MI for
individual modules from a repository:

\[ MI = 171 - 5.2ln(V) - 0.23C - 16.2ln(L) \]

Where $V$ is the Halstead volume, $C$ is the cyclomatic complexity, and $L$ is
the lines of code. MI for a repository is then obtained by averaging over MI for individual modules.

\item We also consider the number of closed issues over a period of time (last 2
years, last 1 year, last 6 months, and last 1 month, with increasing importance
in that order) from the date of evaluation
to determine maintenance activity of a repository.
This is because we want to value current maintenance activity of a repository more.



\end{itemize}

\paragraph{\textbf{Phase 2: Obtaining quality, popularity, and maintainability score of
repositories}}

Once the values of all the measures for all the repositories are obtained, in
the second phase, we calculate scores of all the repositories for all three
categories and then combine them to determine overall score.

Before we calculate the scores of all the repositories, we first normalize
values of all the measures to the range of 0\% to 100\%, with 0\% being the
lowest normalized value and 100\% being the highest. This is performed by
obtaining the lowest and the highest value of every measure and computing the
position of a value within the range of the lowest and the highest value. If we
use $m$ to represent a measure and $V_m$ to represent the set of values for that
measure before normalization (a column in the $R{\times}M$ table), then the
formula for normalization is:

\[ \frac{(v - v_{min}) \times 100}{v_{max} - v_{min}} \forall v \in V_m \forall
m \in M \] 

Normalized values of the measures are then used to compute the quality score
($q_r$), the maintainability score ($x_r$), and the popularity score ($p_r$) of
every repository $r$ as follows.  For maintainability score, $x_r$, we use
different weights for the measures based on their importance towards the score.
Otherwise, we assign equal weight to all the measures for $q_r$ and $p_r$.  We
compute overall score $s_r$ of a repository $r$ as a mean of $q_r$, $x_r$, and
$p_r$.  All of these scores are also normalized to the range of 0\% to 100\%,
with 0\% being the lowest and 100\% being the highest.

\[ q_r = 100 - \frac{n_{cc} + n_{sty} + n_{sl} + n_{sm} + n_{sh}}{5} \]

\[ x_r = \frac{51 \times n_{mi} + 9 \times n_{c2y} + 9 \times n_{c1y} + 9
\times n_{c6m} + 12 \times n_{c1m} + 12 \times n_{cm}}{100} \]

\[ p_r = \frac{n_{ss} + n_{str} + n_{fr}}{3}, \hspace{0.1in}
  s_r = \frac{q_r + x_r + p_r}{3} \]

\section{Evaluation}
\label{section:eval}

As discussed in the introduction section, the objective of this work is to
evaluate if quality of open-source repositories indeed impacts performance of MP
systems. Towards that objective, we choose ControlFlag as a candidate MP
system\footnote{There was no particular reason to select ControlFlag besides the fact that its
source-code is publicly available at
\url{https://github.com/IntelLabs/control-flag}, and its training phase
does not require GPUs but can work well with commodity CPUs.}.

\paragraph{\textbf{A brief background on ControlFlag.}} We provide
background on ControlFlag for the purpose of explaining our results. ControlFlag
is an MP system that aims to find typographical errors in code
automatically. Towards that goal, it uses a self-supervised learning approach
that relies on the dataset of commonly-used programming patterns that is
generated by mining open-source repositories. It formulates the
problem of finding erroneous programming patterns as an anomaly detection problem, where
erroneous patterns are anomalies. ControlFlag clusters mined patterns using
abstract representations, and every pattern along with its statistical
frequency are stored in the dataset. It uses two levels of
abstractions (L1 and L2), with L2 being more abstract than L1.

ControlFlag considers GitHub repositories having at least 100 stars as high
quality and mines patterns from them for training purpose. We conducted
experiments to answer two research questions:
\begin{itemize}
\item \textbf{RQ1:} Does ControlFlag perform poorly on repositories that are
ranked low by {\framework} than high?
\item \textbf{RQ2:} Does ControlFlag perform poorly on low-starred (less than
100 stars) repositories than high-starred (more than 100 stars) repositories?
\end{itemize}

In a sense, both the questions use ControlFlag as a common test to determine if
the source code measures used by {\framework} (in \textbf{RQ1}) and the measures
used by ControlFlag (in \textbf{RQ2}) capture the relative quality of the
repositories. The question that has an affirmative answer would suggest that the
code measures used in that test impact ControlFlag's output. For instance, if
answer to \textbf{RQ1} is affirmative, then it suggests that the code measures
used by {\framework} impact ControlFlag's output. And in the case of
\textbf{RQ2}, an affirmative answer would mean the number of GitHub stars
impacts ControlFlag's output.

We answer these research questions by comparing the datasets obtained from
mining the repositories used in those questions. Specifically, we design two
sets of experiments for our evaluation:


\textbf{Experiment 1.} We designed experiment 1 to answer \textbf{RQ1}. In
particular, we randomly select 500 repositories out of the list of C++
repositories provided in ControlFlag's code repository. (In order to ensure that
ControlFlag's code quality measure does not impact this experiment, we select
500 repositories from a list of repositories having more than 100 stars.)  We
then rank those repositories using {\framework} and divide them into two halves
--- the top half containing 250 repositories having higher rank than the bottom
250 repositories. {\framework} considers the top half as the set of higher-quality repositories
than the bottom half (that is considered as a set of lower-quality
repositories). We then generate training datasets using both the sets and
evaluate quality of those datasets using three tests described below. We denote
the dataset obtained from the top 250 repositories by $D_{t1}$ and the one obtained
from the bottom 250 repositories by $D_{b1}$\footnote{We use suffix $t$ for top
and $b$ for bottom.}.

\textbf{Experiment 2.} The objective of the second experiment is to answer
\textbf{RQ2}.  Towards that end, we use GitHub's REST APIs to obtain two sets of
GitHub repositories having C++ as their primary language: the first set of 250
repositories having more than 100 stars and the second set of 250 repositories
having less than 100 stars. The first set of repositories would be considered
high-quality by ControlFlag (because they have more than 100 stars), while the
second set of repositories would be considered low quality by ControlFlag
(because they have less than 100 stars). Notice, however, that both $D_{t1}$ and
$D_{b1}$ could be used as the dataset obtained from the first set of
repositories as both of them only consume repositories having more than 100
stars. Consequently, we only collected the second set of 250 repositories having
less than 100 stars and generated a dataset (denoted by $D_{b2}$) from it. In
summary, the experiment compares $D_{t1}$ and $D_{b2}$ on the three tests
described below.

We present some of the statistics (number of style errors and security issues)
of the datasets used in the experiments in Table~\ref{table:results}. As
expected, the number of style errors and security issues are low for $D_{t1}$
but high for $D_{b1}$ and $D_{b2}$.


\begin{table*}[!htbp]
\begin{footnotesize}
\begin{tabular}{l||l|l|l|l||p{1cm}|l|l||p{1cm}|l|l}
\hline
\multirow{3}{*}{\textbf{Experiment}} & \multirow{3}{*}{\textbf{Dataset}} &
\multicolumn{3}{c||}{\textbf{Dataset statistics}} & \multicolumn{3}{c||}{\textbf{L1 abstraction level}} &
\multicolumn{3}{c}{\textbf{L2 abstraction level}} \\
\cline{3-11} &
           & \multirow{3}{1cm}{\textbf{Style errors}}
           & \multirow{3}{1cm}{\textbf{Security warnings}}
           & \multirow{3}{1cm}{\textbf{Security errors}}
& \multicolumn{2}{c|}{\textbf{T1}} & \multicolumn{1}{c||}{\textbf{T2}}
                  & \multicolumn{2}{c|}{\textbf{T1}} & \multicolumn{1}{c}{\textbf{T2}} \\
\cline{6-11}
           &  & & &
           & \textbf{Patterns missing} & \textbf{Anomalies} & $c_{degree}$, $c_{stdev}$, $c$
           & \textbf{Patterns missing} & \textbf{Anomalies} & $c_{degree}$, $c_{stdev}$, $c$\\
\hline
\multirow{2}{*}{Experiment 1} & $D_{t1}$ &
5.42 & 0.01 & 0.06 &
\textbf{640} & \textbf{0} & 0.0070, 1.74, -0.73 & \textbf{0} &
\textbf{4} & \textbf{0.17}, 7.28, -6.11 \\
\cline{2-11}
& $D_{b1}$ &
11.94 & 0.06 & 0.27 &
831 & \textbf{0} & 0.0076, \textbf{1.70},\textbf{-0.69} & 0 &
0 & 0.15, \textbf{6.34}, \textbf{-5.19}\\
\hline
\multirow{2}{*}{Experiment 2} & $D_{t1}$ &
5.42 & 0.01 & 0.06 &
640 & \textbf{0} & 0.0070,
\textbf{1.74}, \textbf{-0.73} & \textbf{0} &
\textbf{4} & \textbf{0.17}, 7.28, -6.11 \\
\cline{2-11}
& $D_{b2}$ &
9.49 & 0.06 & 0.17 &
\textbf{591} & \textbf{0} & \textbf{0.0079},1.76, -0.75
& 3 & 3 & 0.15, \textbf{6.57}, \textbf{-5.42}\\
\hline
\end{tabular}
\end{footnotesize}
\caption{Table showing results of the tests \textbf{T1} and \textbf{T2} for
experiment 1 and 2. Dataset statistics are numbers per LoC, summed up for all
the repositories in the dataset. Bold values in columns show
the best result for that test across both the experiments. (Results of test \textbf{T3} are
not presented in the table for the purpose of space. They are rather discussed
in text.)}
\vspace{-0.12in}
\label{table:results}
\end{table*}


%
%

\begin{figure}
\includegraphics[width=0.45\textwidth]{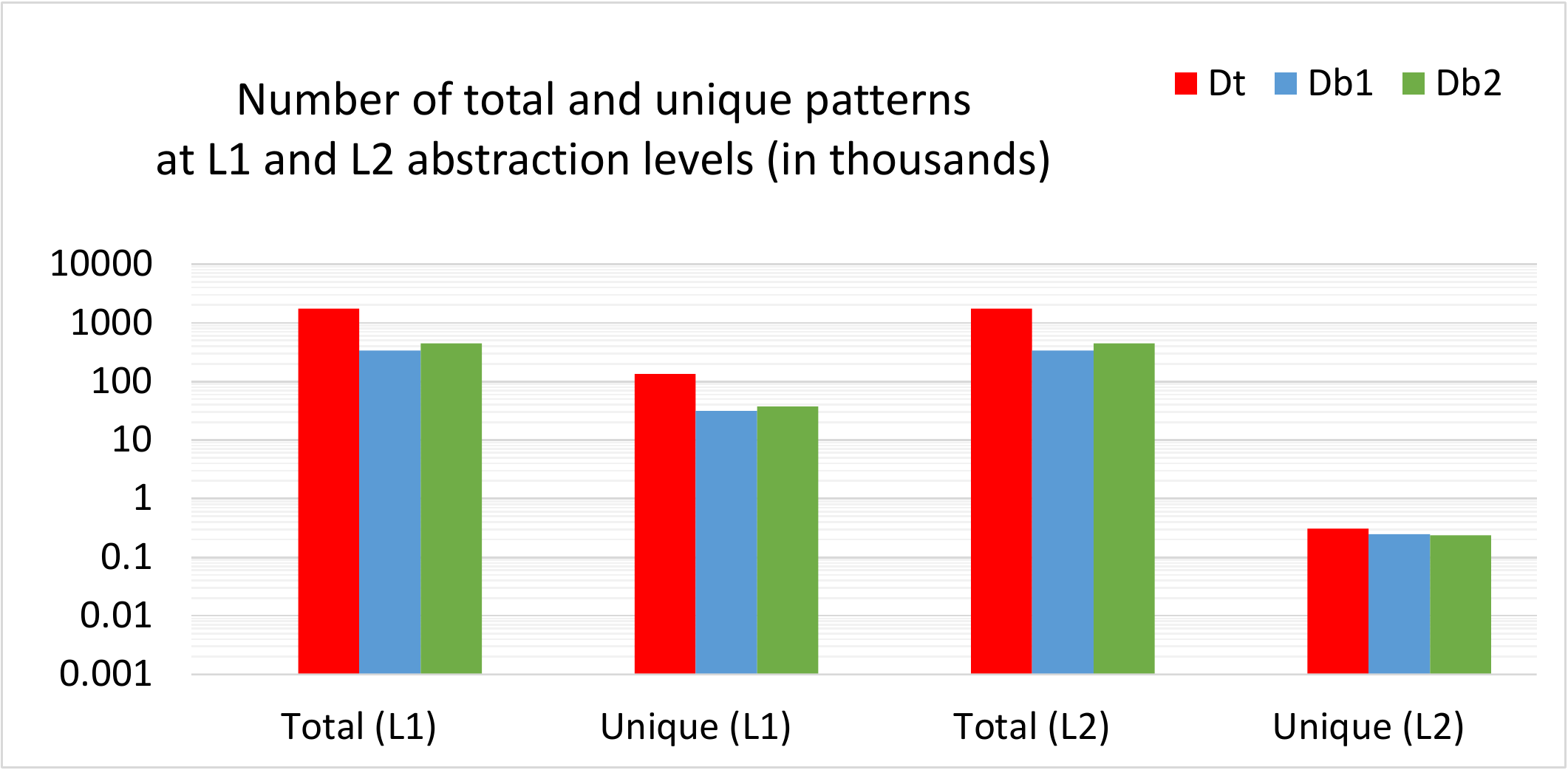}
\caption{Diversity and coverage of patterns in the datasets used in different
experiments}
\label{fig:dataset_coverage}
\vspace{-0.12in}
\end{figure}

Below we describe tests to compare ControlFlag's training datasets.

\textbf{T1}: In the first test, we apply the training datasets to
train ControlFlag individually and then apply it to find programming errors in
\texttt{https://github.com/microsoft/DirectXTex}\footnote{a randomly-selected
candidate repository having C++ as its primary language and more than 100 GitHub
stars.}.

\textbf{T2}: In the second test, we visualize ControlFlag's datasets as graphs
and use graph analysis techniques to ensure that the datasets are not corrupted
by adversarial attacks. Such attacks would degrade quality of the datasets.

In ControlFlag, the training dataset consists of a set of programming patterns
that are mined from repositories. ControlFlag uses a decision-tree based
approach to cluster the mined patterns. Let's say that the clustered training
data, obtained from $K$ open-source repositories, consists of $J$ distinct
patterns, where every pattern $p$ has $n$ occurrences. In other words, the
training dataset consists of $J$ tuples, where every tuple is of the form: ($p$,
$n$, $d$), where $p$ is a pattern, $n$ is the number of occurrences of that
pattern in the training dataset, and $d$ is the number of repositories
contributing to pattern $p$. Then, the training dataset can be visualized as a
directed bipartite graph, where edges originate in the repositories and
terminate in the distinct patterns. Every edges also carries the weight of the
number of occurrences $n_{kj}$, representing the contribution of $k$th
repository to $j$th pattern.

Given this formulation, the second test measures the \emph{confidence} ($c$) in the
training dataset using a consensus based approach. Intuitively, we have more
confidence in the training dataset when:
\begin{itemize}

\item The majority, if not all, patterns receive contributions from several
repositories, if not all.  This is measured as the degree of connectivity in the
graph and denoted by $c_{degree}$.  Intuitively, $c_{degree}$ ensures that the
patterns are ``peer-reviewed'' by many repositories (or peers).

\item When every pattern receives similar, if not exactly same, contribution
from every repository (measured as $c_{stdev}$).  Intuitively, $c_{stdev}$
measures the deviation in the training dataset in terms of contributions of
various repositories to different patterns and ensures that no one repository
skews the confidence in a pattern by contributing heavily to it.

\end{itemize}

Both of the above criteria ensure that a single repository does not skew
training dataset in ControlFlag by contributing heavily than other repositories.
Such an adversarial case would degrade the quality of the dataset. Note however
that a collusion attack is still possible wherein multiple repositories
collectively contribute patterns in malicious manner to degrade quality of the
dataset.

\[ c = c_{degree} + (1 - c_{stddev}) \]
\[ c_{degree} = \frac{\sum_{i=1}^{J}\frac{indegree(p_{i})}{K}}{J}, \hspace{0.1in}
   c_{stdev} = \frac{\sum_{i=1}^{J}\sigma(\{n_{1i}, n_{2i}, .., n_{Ki}\})}{J} \]

High value of $c_{degree}$ shows high confidence, while high value of
$c_{stdev}$ shows low confidence in the dataset.

\textbf{T3}: The third test is a qualitative one. We check training datasets for
examples of good and bad programming patterns to determine if good-quality
repositories indeed use good programming practices.

%

Table~\ref{table:results} shows the comparison of three training datasets across
\textbf{T1} and \textbf{T2} and L1 and L2 abstraction levels.

\paragraph{\textbf{Result analysis.}} The results answer \textbf{RQ1} in
affirmative and \textbf{RQ2} in negative. Specifically, in experiment 1, dataset
$D_{t1}$ wins test \textbf{T1} for both L1 and L2 abstraction levels. It,
however, loses to $D_{b1}$ on test \textbf{T2}. We consider test \textbf{T1} as
more authoritative as it is administered by ControlFlag itself --- \textbf{T2}
uses our graph based formulation, which is independent of ControlFlag.  Test
\textbf{T1} reports the number of patterns from the evaluation repository that
are not found in the given dataset. In other words, it compares the
\emph{coverage} of the patterns in the given dataset --- less number of missing
patterns indicates better coverage.  Finally, recall that $D_{t1}$ is marked as
higher-quality than $D_{b1}$ by {\framework}.

Experiment 2, on the other hand, does not have a clear winner. Dataset $D_{t1}$,
marked as high-quality based on ControlFlag measures, wins test
\textbf{T1} for L2 abstraction level, while it loses test \textbf{T1} for L1
abstraction level.  Similarly, dataset $D_{b2}$, which is marked as low-quality
based on ControlFlag measures, wins test \textbf{T2} for L1 abstraction level,
while it loses test \textbf{T2} for L2 abstraction level. Recall that in this
experiment, dataset $D_{t1}$ is marked as high-quality while dataset $D_{b2}$ is
marked as low-quality by ControlFlag.

The anomalies for test \textbf{T1} indicate possible programming errors in code.
As such, $D_{t1}$ finds 4 anomalies in the evaluation repository and performs
better than others. In summary, when we compare the datasets directly --- a
standalone statistical test without the evaluation repository --- using the
number of total and unique patterns (as shown in
Figure~\ref{fig:dataset_coverage}), then $D_{t1}$ comes out to be winner in
experiment 1 and 2.


Across both the experiments, the test \textbf{T2} marks $D_{b1}$ as an overall
winner on $c$, but different datasets win on the basis of $c_{degree}$ and
$c_{stdev}$. This indicates that some datasets have high degree of peer-review
activity (e.g., 0.17 value of $c_{degree}$ for $D_{t1}$) and
high confidence, while some datasets have skewed contributions of patterns
(e.g., 7.28 value of $c_{stdev}$ for $D_{t1}$) and low confidence.

The authors of the paper performed qualitative inspection of all three datasets
(in test \textbf{T3}) and found that all of them contained instances of abnormal
patterns that indicate bad programming practices\footnote{One common bad
programming practice we found is to perform assignment inside a conditional
statement without parenthesis (e.g., \texttt{if (x = 5)}). Good programming
practice is to use parenthesis around such statements (e.g., \texttt{if ((x =
5))}). We consider expressions as bad programming practices if GCC warns for
those expressions~\cite{gcc_warning} (e.g., \texttt{-Wparentheses} for the
assignment without parenthesis case) and by referring to standard style guides
such as Linux kernel~\cite{kernel_styleguide}.}. Even dataset $D_t$ contained confusing
programming expressions, such as \texttt{if (x < string\_literal)} and redundant
expressions such as \texttt{if (false \&\& x)}, indicating that not all
top-quality repositories are using good programming practices.

\paragraph{\textbf{Overall thoughts and future work.}} Overall, we find that the
results show some correlation between the quality measures used by {\framework}
and the output of ControlFlag. In other words, selecting lower-quality
repositories marked by {\framework} would degrade ControlFlag's performance. We
also find that the results show negative correlation between the quality
measures used by ControlFlag and its output. Precisely, selecting repositories
having less than 100 stars does not necessarily impact ControlFlag's output.
This point corroborates the known observation that the number of GitHub stars is
not a quality measure but is a popularity measure.

We nonetheless acknowledge that the results also suggest some future work. To be
precise, not all the code measures used in {\framework} seems to affect
ControlFlag. Specifically, ControlFlag can handle source code having
compilation issues, suggesting that the absence of compilation issues is not its
quality measure. This observation that quality measures that actually influence
the output of ControlFlag could be tuned versions of the typical quality measures
considered in software engineering literature raises some interesting questions.
It is not clear if similar observations would apply to other MP systems.  We
plan to answer these questions in the future.

\paragraph{\textbf{Reproducibility.}} We have published all the data and scripts
used in both the experiments publicly under {\framework} at
\url{https://github.com/nirhasabnis/gitrank/tree/main/case_study/evaluate_MP_systems}.
{\framework} is also available publicly at
\url{https://github.com/nirhasabnis/gitrank}. ControlFlag used in our evaluation
is available at \url{https://github.com/IntelLabs/control-flag}.

\section{Related Work}
\label{section:relwork}

To the best of our knowledge, we are not aware of any existing work that
evaluates impact of quality of open-source repositories on MP systems.
Nevertheless, several existing efforts have developed code metrics and models to
evaluate open-source projects on quality, maintainability, popularity, among
other criteria~\cite{Jarczyk:2014, ludwig:2017, ludwig:2019:cbr,
munaiah:2017:curating, samoladas:2008:sqooss, stamelos:2002:codequality,
spinellis:2009:sqooss1}. We briefly summarize some of them below. Stamelos et
al.~\cite{stamelos:2002:codequality} compare open-source software development
model with the closed-source model by considering structural quality of code and
measuring it for code developed using open-source style development.  Spinellis
et al.~\cite{spinellis:2009:sqooss1}, on the other hand, apply SQO-OSS
platform~\cite{samoladas:2008:sqooss} and combine process and product metrics to
evaluate quality aspects of open-source software.
Jarczyk et
al.~\cite{Jarczyk:2014}, on the other hand, evaluate correlation between quality
of GitHub projects and characteristics of their team members, thereby analyzing
the social aspect of open-source software development. Specifically, they
develop metrics reflecting project's popularity and quality of support offered
by its team members and apply statistical regression techniques to analyze their
influence on project quality.

\section{Threats to validity}

In this preliminary paper, we selected some of the commonly-used source code measures
in software engineering community. As we mentioned previously, our
selection is currently ad-hoc. Consequently, it is possible that we could have missed some
measures that could lead to better or worse results.  Nonetheless, the existing
results stand for the given set of selected measures.  Another related but
different point is about multicollinearity between different measures.
Specifically, if multiple measures are correlated with each other by any mean,
then they would have higher influence over the results than other measures. We
consider the overall question of systematic selection of measures as a future
work.

Although, the description and design of our approach is generic and applicable
to any MP system, our current evaluation is restricted to ControlFlag. As such,
the conclusions would be specific to ControlFlag also (in other words,
generalizing them would be incorrect.) Moreover, we used code repositories
hosted on GitHub for our study (as ControlFlag only supports GitHub
repositories), nonetheless several other hosting platforms exist.

\section{Conclusion}
\label{section:conclusion}

Machine programming (MP) is an emerging field that combines probabilistic
approaches (such as artificial intelligence) and deterministic approaches (such
as formal methods) to solve problems in software engineering and systems.
Several of the existing machine programming systems rely on open-source code
without considering its quality, or use atypical quality measures (than the ones
typically used in software engineering community).

In this preliminary study, we developed a framework to rank open-source
repositories on quality, maintainability, and popularity, and applied it to
generate sets of repositories of different quality levels. We then used those
repositories to generate training datasets for ControlFlag and analyze their
impact on ControlFlag's performance. Our results so far indicate that
ControlFlag's quality measure --- the number of GitHub stars --- does not
correlate strongly with its performance. On the other hand, the results also
indicate that the code quality measures typically used in software engineering
have higher correlation with ControlFlag's performance. We nonetheless observe
that certain code quality measures used in {\framework} does not affect
ControlFlag, and we may need to finetune such measures based on an MP system.
The results also raise several questions for other MP systems to be addressed in
the future.

\balance
\bibliography{main}
\bibliographystyle{ACM-Reference-Format}

\end{document}